\documentclass[12pt]{iopart}

\usepackage{iopams}  
\usepackage{color}
\usepackage{psfrag} 
\usepackage{graphicx}
\usepackage{amssymb}

\def\bx{{\bf x}}

\begin{document}

\title{Variational approach for Bose-Einstein condensates in strongly disordered traps.}
%with dipolar interaction.}

\author{G. M. Falco}

\address{Institut f\"ur Theoretische Physik, Universit\"at zu
K\"oln, Z{\"u}lpicher Str. 77, D-50937 K\"oln, Germany}
\ead{falco@thp.uni-koeln.de}

\begin{abstract}
Recently, Nattermann and Pokrovsky [PRL {\bf{100}}, 060402 (2008)]
have proposed a scaling approach for studying
Bose-Einstein condensates in strongly disordered traps. 
In this paper we implement their scaling argument
in the framework of the variational method
for solving the time dependent Gross-Pitaevskii equation.
We consider atomic gases with both short range $s$-wave interaction and 
long range anisotropic dipolar interaction. 
The theory is addressed to
%novel quantitative predictions about 
the regime of strong disorder and weak interactions
where the physics is dominated by the collective pinning due to the disorder. 
The phenomenon of condensate fragmentation in dipolar gases is also analyzed.

\end{abstract}
%%In a recent work (PRL {\bf{100}}, 060402 (2008))

 
\maketitle

\section*{Introduction}

Motivated by the recent progresses in ultracold alkali atoms in disordered traps
\cite{Aspect08,Fallani08,Fortagh07}, 
Nattermann and Pokrovsky proposed \cite{Nattermann08} 
a semiquantitative approach for Bose-Einstein condensates in strong random potentials where standard 
perturbative methods are expected to fail.
Their analysis is based on the evaluation of
the mean-field energy 
due to the 
fluctuations of the random potential according
to a scaling argument                 
introduced by Larkin~\cite{Larkin70} 
for studying the effect of defects in flux line lattices.
%In the case of a Bose-Einstein condensate the kinetic energy replaces the elastic energy.
The scaling approximation requires that the correlation length of the disorder is much shorter than the correlation length
of the liquid.
In this paper, we implement the method of Nattermann and Pokrovsky
in the framework of the time dependent variational method 
of Perez-Garcia {\it et al.}~\cite{Zoller97}. As a result, we obtain a simple set of equations, which describe the 
equilibrium and the low-energy dynamics
at zero temperature of a Bose-Einstein condensate in a strong random potential.
Beside the oscillator and the scattering length,
a new quantity enters into the problem, namely the Larkin length ${\mathcal L}$ associated with the collective pinning 
of the condensate.

The time-dependent variational method represents 
a very good approximation of the Gross-Pitaevskii equation \cite{Kimura02,Salasnich04}.
Moreover, it can interpolate quite successfully from the low density regime to the strong coupling
Thomas-Fermi gas. The present extension to disordered traps, however, presents some restrictions.
Since the fluctuating center of attraction of the disorder may not coincide with the center of the harmonic trap,
the model is fully consistent only when either the harmonic trap or the disorder 
is responsible for the localization of the condensate.
Moreover, the Gaussian variational ansatz
out-rules from the outset the solutions describing the multi-fragmented state of the condensate~\cite{Nattermann08}.
Despite these limitations, %we will show that we can derive 
novel quantitative predictions can be obtained with regard to 
the disorder dominated regime. 
In this limit the condensate becomes non-superfluid and
%dominated by the Larkin length.
%which has 
%not been explored in the experiments yet, 
is characterized by a 
{\it generalized} correlation length~\cite{heal} %$\xi_{\rm heal}$
larger than the size of the cloud.
%and has not been explored in the experiments yet.

The paper is organized as follows. In Section \ref{sec1a} 
we introduce the general formulation of the problem.
We determine the size of the  
ground state at equilibrium 
and we calculate the expressions for the frequencies of the low-lying excitations. 
In Section \ref{sec1b} we discuss in detail the $3$D gas in presence of a strong delta-correlated disorder.
We show that, reducing the atomic scattering length by means of Feshbach resonance techniques,  
the predictions of the theory for the single connected fragment might be tested experimentally.
In Section \ref{sec1c} low-dimensional gases are considered.
Section \ref{sec1d} contains some considerations on the applicability of the method 
in the case of disorder with finite correlation length. 
In Section \ref{sec2a} we introduce a long-range anisotropic dipolar interaction and we study the stability 
of the single connected fragment. The nature of fragmentation in presence of dipolar interaction is 
analyzed in Section  \ref{sec2b}. The static on-average characteristic properties of the fragments are determined
at the level of the Imry-Ma level of approximation~\cite{Imry75}. 

\section{Time dependent variational method with disorder}

\subsection{Variational equations}
\label{sec1a}

We begin considering the Hamiltonian of a gas of trapped bosons interacting through $s$-wave scattering 
\begin{eqnarray}
\label{eq:hamiltonian}
\hat{H}=\int d^3x \hat\Psi^{\dagger}\left(
-\frac{\hbar^2}{2m}\nabla^2+V_{trap}(\bx)+U\left(\bx\right)+\frac{4\pi\hbar^2 a }{m}\hat\Psi^{\dagger}\hat\Psi
\right)\hat\Psi,
\end{eqnarray}
where $a$ is the $s$-wave scattering length and the harmonic trapping potential is
$V_{trap}(\bx)=\sum_{i=x,y,z}m\omega_i^2 R_i^2/2$.
The disorder potential $U(\bx)$ is chosen to be Gaussian distributed characterized by the average values
$\langle U\rangle=0$ and $\langle U(\bx )U({\bx}')\rangle=\left(\kappa^2/b^3\right) K_{b}\left(\bx-\bx'\right)$, 
where $b$ denotes the correlation length of the disorder. We will assume
that $K_{b}(\bx)$ is a smeared out $\delta$-function $K_{b}(\bx)={\rm e}^{-\bx^2/b^2}
/{(2\pi)^{3/2}}$. 

The variational calculation follows the same outline
of the clean case. Therefore, in what follows, we refer the reader to~\cite{Zoller97}
for the details of the derivation.
We consider the normalized variational wave-function 
in the case of a $3$-D fully anisotropic configuration
\begin{eqnarray}
\label{eq:Zoller_Ansatz}
\psi\left(\bx,t\right)=
\frac{N^{1/2}}{\pi^{3/4} {\tilde{R}^{\frac{3}{2}}(t)}}
{e}^{-\sum_i^{3}\left[\frac{1}{2 {R_i^2(t)}}+i B_i(t)\right]x_i^2},
\end{eqnarray}
where we have defined the geometric average $\tilde{R}=\left[
R_x(t) R_y(t) R_z(t)\right]^{1/3}$ and the variables
$R_i$ and $B_i$ are time dependent variational parameters.
{\color{black}The $R_i$'s are related to size of the system while the imaginary
width $i B_i$ of the Gaussian ansatz is necessary to include the dynamics into the 
variational procedure. Without this latter
the minimum principle would just lead to the condition of static equilibrium.
 }
 
Physically, the true ground state of the stationary Gross-Pitaevskii
equation in the presence of interactions and disorder 
is expected to deviate strongly from the ground state of the noninteracting system. Nevertheless,
for a clean system, 
it has been shown that
the dynamical trial wave-function~(\ref{eq:Zoller_Ansatz}) 
is a very good approximation of the Gross-Pitaevskii at finite $N$ \cite{Zoller97,Kimura02,Salasnich04}. 
%it can
%interpolate between the noninteracting and the strong interacting system \cite{Zoller97,Kimura02,Salasnich04}. 
%Therefore, it seems reasonable to apply it to the low-energy oscillations also in the presence of disorder.
Inserting the trial-wave function 
in the Lagrangian relative to the Hamiltonian of~(\ref{eq:hamiltonian})
we find
\begin{eqnarray}
\label{eq:Var_Lagr}
L=L_{0}+\frac{\kappa}{\pi^{3/4}}\frac{N}{{\prod}_{i=1}^3
\left[2 R_i^2(t) +b^2\right]^{1/4}}
\end{eqnarray}
where $L_{0}$ is the well known contribution in absence of disorder~\cite{Zoller97}.
The new term represents the on-average contribution of the disorder fluctuations 
according to the scaling argument of Nattermann and Pokrovsky~\cite{Nattermann08,Larkin70}.
It introduces a new relevant length scale,
namely the Larkin length 
${\mathcal L}=({\color{black}2^{3/2}}\pi^{3/2}\hbar^4/m^2\kappa^2)$,
associated with the {\color{black} pinning energy due to the disorder}~\cite{meanfreepath}. 
{\color{black}
The length scale ${\mathcal L}$ was introduced by Larkin~\cite{Larkin70} 
in connection with the onset of collective pinning of vortex lines in type-II superconductors.
In the case of a trapped gas, 
the analogous of the (delocalizing) elastic
energy relative to the lattice distortion 
corresponds to the kinetic energy of the atoms.
}
Since
the scaling argument requires to have {\color{black}both ${\cal L}$ and $b$ much smaller than the generalized
healing length $\xi_{\rm heal}$~\cite{heal}, {\it i.e.}}
${\mathcal L}\ll \xi_{\rm heal}{\color{black} }$ and $b\ll \xi_{\rm heal}$,
the validity of ~(\ref{eq:Var_Lagr})
is in general restricted to the quantum limit $b\leq {\mathcal L}$.
The classical limit $b\gg {\mathcal L}$, when many levels occupy the typical well,
will be briefly discussed separately.

In what follows it is convenient to rescale the quantities in unities 
of the harmonic oscillator such as $r_i=R_i/{\ell}$
and $({\tilde b}/r_i)\equiv(b/\ell)(\ell/R_i)$ where $\ell=\sqrt{\hbar/m \omega}$ is the harmonic oscillator length.
The anisotropy of the external trapping is taken into account by setting $\omega_i=\lambda_i \omega$.
Without loss of generality we will consider a trap with cylindric symmetry with anisotropy factors $\lambda_x=\lambda_y=1$ 
and $\lambda_z=\lambda$. In absence of disorder this implies that the angular momentum
along the $z$ axis is conserved and we can label the modes by the azimuthal angular quantum numbers $m$.

{\color{black} 
Next, we derive the Euler-Lagrange equations for 
the variables $R_i(t)$ and $B_i(t)$ relative to the Lagrangian~(\ref{eq:Var_Lagr}). 
Then, eliminating the $B_i$'s variables, we obtain
a closed system of differential equations for the $R_i$'s. These equations
can be viewed as describing the motion of a point-like particle in an effective potential. 
The position 
of equilibrium at rest of the particle
is determined by the minima of the effective potential.
Using rescaled unities $r_{0x}=r_{0y}$ and $r_{0z}$,
this leads ultimately
to the conditions 
%~\cite{Zoller97} we find the equilibrium conditions 
}
\begin{eqnarray}
\label{eq:equil_cond_x}
r_{0x}^4+\gamma
\frac{r_{0x}}{\sqrt{r_{0z}}}
{f_{x}}
{\prod_{j=1}^3 f_{j}^{1/4}}
=1 + \alpha\frac{1}{r_{0z}}\\
\label{eq:equil_cond_z}
\lambda^2 r_{0z}^4+
\gamma\frac{r_{0z}^{3/2}}{r_{0x}}
{f_{z}}
{\prod_{j=1}^3 f_{j}^{1/4}}
=1+ \alpha\frac{ ~r_{0z}}{r_{0x}^2},
% \label{eq:equil_cond}
\end{eqnarray}
where we have defined the functions {\small
$f_{i}\left(r_{0i}\right)\equiv {\color{black}2}\left[2+\left({\tilde b}/{r_{0i}}\right)^2\right]^{-1}$} and
{\small$g_{i}\left(r_{0i}\right)\equiv \left({{\tilde b}}/{r_{0i}}\right)^2 {f_{i}}/{\color{black}2}$,
}
%\begin{eqnarray}
%f_{r_{0i}}\equiv \frac{1}{2+\left({{\tilde b}}/{r_{0i}}\right)^2}\\
%g_{r_{0i}}\equiv \left({{\tilde b}}/{r_{0i}}\right)^2 f_{r_{0i}}/2,
%\label{eq:gfunct}
%\end{eqnarray}
and we have introduced the Thomas-Fermi parameter
$\alpha=\sqrt{{2}/{\pi}}\left({N a}/{\ell}\right)$ together with the disorder strength parameter
$\gamma=\sqrt{{\ell}/{\mathcal L}}$. 
%in terms of the $3d$ Larkin length ${\mathcal L}=({\color{red}2^{3/2}}\pi^{3/2}\hbar^4/m^2\kappa^2)$.
In these two latter equations, the l.h.s. describes the confinement 
due to the trap and to the disorder. In
the r.h.s. this effect is counterbalanced by the 
kinetic energy and the repulsive 
mean-field interactions.   
The limit $\mathcal L\rightarrow \infty$ corresponds to the theory for a clean system~\cite{Zoller97}. 
%which
%is known to be a valid approximation at finite $N$ and in the limit of large $N$ it reproduces 
%the analytical results found by Stringari \cite{Stringari97} 
%for the monopole and quadrupole
%modes 
%in the hydrodynamic regime
%~\cite{Zoller97,Stringari97} .
%
%In the case of a $3d$ isotropic trap in the Thomas-Fermi limit,
%where the equilibrium conditions can be approximated as
%$\alpha\approx r_{0x}^4 r_{0z}$ and
%${r_{0x}^2}\approx\lambda^2 r_{0z}^2$,
%one has $\omega_a=\omega_c=\sqrt{2}\omega$
%and $\omega_b=\sqrt{5}\omega$. 

{\color{black}The low-lying excitations of the system can be calculated by means of a harmonic
expansion $r_{i}(t)=r_{0i}+\delta r_i(t)$
around the equilibrium position~(\ref{eq:equil_cond_x}) and (\ref{eq:equil_cond_z}).
Making a dynamical ansatz for the $\delta r_i(t)$, the calculus of the frequencies
for the small oscillations
is reduced to the calculation of the eigenvalues
of a $3\times 3$
real symmetric matrix ${\rm A}_{ij}$}
whose elements are
%{\tiny$\left(\begin{array}{ccc}
%a & b & c\\
%b & a & c
%c & c & d
%\end{array}\right)$}.
%%%%%%%%%%%%%%%%%%%%%%%%%%%%%%%%%%% Formula analitica degli autovalori
%\end{eqnarray}
%by means of 
%\begin{eqnarray}
%\label{eq:genfreq}
%& 
%{\small $\omega_a=\omega\sqrt{a-b}$},
%\\
%& 
%{\small$\omega_{b,c}=\omega\frac{1}{\sqrt 2}\left[(a+b+d)\pm\sqrt{(a+b-d)^2+8 c^2}\right]^{1/2}$}.
%\end{eqnarray}
%%%%%%%%%%%%%%%%%%%%%%%%%%%%%%%%%%%%%%%%%%%%%%%%%%%%%%%%%%%%%%%%%%%%%%%%%%%%%%%%%%%%%%%%%%%%%%%%%%%%%%%%%%
\begin{eqnarray}
\label{eq:matrx_el_gen}
{\rm A}_{11}=&{\rm A}_{22}=1+\frac{3}{r_{0x}^4}+\frac{2\alpha}{{r_{0x}^4 r_{0z}}}-
\frac{3\gamma}{2{r_{0x}^3 \sqrt{r_{0z}}}}
 f^{3/2}_{x}f^{1/4}_{z}\left\{ 1-\frac{5}{3}g_{x}\right\}
 \nonumber\\
{\rm A}_{12}=&\frac{\alpha}{{r_{0x}^4 r_{0z}}}-
 \frac{\gamma}{2{r_{0x}^3 \sqrt{r_{0z}}}}
 f^{3/2}_{x}f^{1/4}_{z}\left\{ 1-g_{x}\right\}
 \nonumber\\
{\rm A}_{13}=&\frac{\alpha}{{r_{0x}^3 r_{0z}^2}}  
 -\frac{\gamma}{2{r_{0x}^2 {r_{0z}}^{3/2}}}
 f^{3/2}_{x}f^{1/4}_{z}\left\{ 1-g_{z}\right\} 
 \nonumber\\
{\rm A}_{33}=&\lambda^2+\frac{3}{r_{0z}^4}+
  \frac{2\alpha}{{r_{0x}^2 r_{0z}^3}}
  -
 \frac{3\gamma}{2{r_{0x} {r_{0z}^{5/2}}}}
 f^{1/2}_{x}f^{5/4}_{z}\left\{ 1-\frac{5}{3}g_{z}\right\}.
% \nonumber\\
%e&=\frac{\alpha}{{r_{0x}^3 r_{0z}^2}} 
% -\frac{\gamma_0}{2{r_{0x}^2 {r_{0z}^{3/2}}}}
% f^{1/2}_{r_{0x}}f^{5/4}_{r_{0z}}\left\{ 1-g_{r_{0x}}\right\},
\end{eqnarray}
We denote the eigenvalues as $\omega_{\rm a,b,c}$. In absence of disorder $\omega_{\rm a,c}$ corresponds to the frequencies of the 
quadrupole modes with quantum numbers $n=0$ and $l=2$, 
while $\omega_{\rm b}$ 
is the frequency of the monopole mode with $n=1$, $l=0$. 

It is important to remember that the scaling argument leading to~(\ref{eq:Var_Lagr}) applies 
under the strong disorder condition $\ell \gg{\mathcal L}$. 
Since the center of the attractive domain formed by the disorder may not coincide with the center of the harmonic trap,
the variational ground state is exact only when
the system is trapped either by the static fluctuations of the random potential or by the harmonic trap.
{\color{black}
In principle, one could try to extend the ansatz~(\ref{eq:Zoller_Ansatz}) including an
additional variational parameter which describes the center of the condensate. However, the
disorder term in~(\ref{eq:Var_Lagr}) would not bring any new contributions to the equation
of motion of the new variable with respect to that of the clean model~\cite{Zoller97}. This shows that
the interplay between the two different confining mechanisms cannot be incorporated
in the theory, which is a direct consequence of the {\it on-average} nature of the 
Larkin energy considered in~(\ref{eq:Var_Lagr})}.
Therefore, the crossover region has to be considered {\color{black} at best} as an extrapolation.

Moreover, our ansatz for the wave-function cannot describe a multi-fragmented state,
which is expected to occur in a wide range of parameters~\cite{Nattermann08,frgm}. 
In this case our solution may occur as a metastable state \cite{Nattermann08}. 
%In the fragmented state definite values of the frequencies are not expected.  
Due to these limitations, in the next sections we will focus our analysis mainly to 
the single fragment non-superfluid regime 
where the cloud is trapped by the disorder
and the physics is dominated by the Larkin length.

\begin{figure}
\vspace{2cm}
\includegraphics[width=145mm,height=40mm]{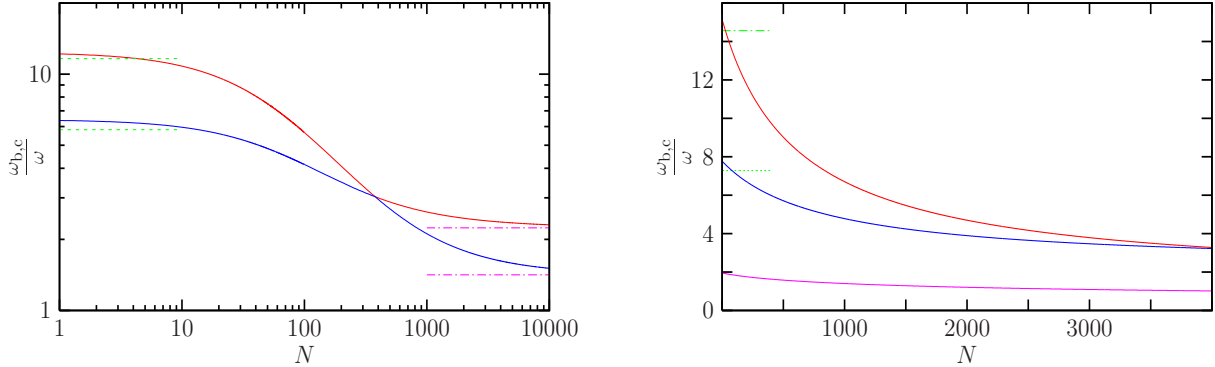}
  \caption{ (a)
  Frequency of the breathing mode $\omega_{\rm b}$ (red solid line) and of the quadrupole ${\rm c}$-mode 
  (blue solid line) as a function of the particle number.
  The upper horizontal green dashed lines describe the analytic results of~(\ref{eq:mode_b_strg_dis_b_0_smN})-
  (\ref{eq:mode_c_strg_dis_b_0_smN})
  while the lower dashed magenta correspond to the Thomas-Fermi limit $\omega_b=\sqrt{5}\omega$
  and $\omega_{\rm c}=\sqrt{2}\omega$. 
  We have considered a sample of $^{87}$Rb atoms confined in an isotropic trap with frequency $\omega=2\pi 50$ Hz
  and 
  with scattering length $a=50 a_0$, where  $a_0$ is the atomic Bohr radius.
  We have considered delta correlated disorder with
  %The external random potential is
  %characterized by zero correlation length and 
  ${\mathcal L}=200 a$ which amounts to 
  $\ell/{\mathcal L}\approx 2.9$. 
  (b) Same frequencies as in (a) %$\omega_b$, $\omega_c$ 
  and the ratio $\omega_b/\omega_c$ (magenta line), 
 % function of the number of
 % particles, 
 for a very shallow trap with $\omega=2\pi 10$ Hz and for a very small value of the scattering length $a=10a_0$.
 % The dashed lines correspond to the analitycal solutions 
 % of~(\ref{eq:mode_b_strg_dis_b_0_smN})-(\ref{eq:mode_c_strg_dis_b_0_smN}). 
 % We have considered delta correlated disorder with 
  The Larkin length is $\mathcal{L}= 2000a$ which means 
  $\ell/\mathcal{L}\approx 3.2$. %Note that at $N=1000$ the density of the gas of $\sim 2.5 \times 10^{14} cm^{-3}$. 
} 
  \label{figureA}
\end{figure}

\subsection{Strong disorder with zero correlation length}
\label{sec1b}

In the limit of very short correlated disorder $b\leq{\mathcal L}$ the corrections due to finite
correlation length are very small. Therefore, as first approximation, we can consider 
a delta correlated disorder, and put $b=0$, {\color{black}$f_i=1$} and $g_i=0$ in~(\ref{eq:Var_Lagr})-(\ref{eq:matrx_el_gen}). 
{\color{black} Moreover, in this limit and for low energy oscillations,
the wave-length of the modes is much larger than the distance over which the disorder varies.
Therefore, the oscillations can be considered self-averaging and we expect the values of the frequencies
to be independent of the specific realization.}

At low densities and strong disorder, when both the radial
and the axial oscillator 
lengths of the traps are larger than the Larkin length, 
the gas is confined mainly by the random potential and the cloud is spherically symmetric.
The equilibrium conditions in~(\ref{eq:equil_cond_x}) and  ~(\ref{eq:equil_cond_z}), which determines the size of the system, 
can be approximated by
%\begin{eqnarray}
%\label{eq:equil_cond_b_zero_strg_dis}
$\gamma\sqrt{r_0}\simeq1+\alpha/r_0$.
%\end{eqnarray} 

\noindent In the limit $N\ll ({\mathcal L/a})$ the interactions can be neglected, the size of the cloud is 
$R\simeq {\mathcal L}$ \cite{mettinota} and we find
\begin{eqnarray}
\label{eq:mode_b_strg_dis_b_0_smN}
\omega_{a}=\omega_b=&c_0\omega\left({\ell}/{\mathcal L}\right)^2\\
\label{eq:mode_c_strg_dis_b_0_smN}
\omega_{c}=\left(c_0/2\right)&\omega\left({\ell}/{\mathcal L}\right)^2,
\end{eqnarray}
where the constant $c_0$ depends on the normalization of the variational trial wave-function we have chosen.
Nevertheless, although 
the exact physical value of the constant $c_0$ cannot be rigorously determined by our approach,
%because depends on the normalization of the variational trial wave-function we have chosen.
the ratio $\omega_{b}/\omega_{c}=2$ of the quadrupole and the monopole does not depend on it. 
For larger number of particles  
${\mathcal L}\ll N a\ll \left({\ell}/{\mathcal L}
\right)^{5/7}\ell$ the size of the system is determined by the competition 
between disorder and interaction and we find 
\begin{eqnarray}
\label{eq:mode_a_strg_dis_b_0_TF}
&\omega_a=\omega_c\propto \omega \frac{\ell^2}{(Na)^{7/6}{\mathcal L}^{5/6}}
\left(\frac{{\mathcal L}}{Na}\right)^{1/6}
\\
\label{eq:mode_b_strg_dis_b_0_TF}
&\omega_b\propto \omega \frac{\ell^2}{(Na)^{7/6}{\mathcal L}^{5/6}}.
\end{eqnarray}
In this interval the generalized healing length $\xi_{\rm}$  reaches at some point the size of the system
which is of the order of $(N a)^{2/3}{\mathcal L}^{1/3}$. Moreover,
the two curves exhibit  an ``avoided crossing-like" feature.
For even larger number of particles   
the external harmonic potential dominates over the disorder 
and the frequencies approach the well-known 
Thomas-Fermi  \cite{Stringari96}.
The crossover between the different regimes of the values of the quadrupole and monopole modes
is illustrated in figure~\ref{figureA}{a}
for some typical experimental parameters. %The curves exhibit  an ``avoided crossing-like" feature.

At $N\gg \left({\ell}/{\mathcal L}\right)^{5/7}(\ell/a)$ the crossover to the Thomas-Fermi regime 
of the clean theory occurs~\cite{Nattermann08}. 
Deep into the Thomas-Fermi regime 
the system is superfluid with $\xi_{\rm heal}\ll {\mathcal L}$.
{\color{black}The critical number of particles 
$N_{\rm c}$ where transition to superfluid occurs
can be estimated by equating the Larkin length
with the superfluid healing length obtained from eqs.~(\ref{eq:equil_cond_x})-(\ref{eq:equil_cond_z})
and the definition given in~\cite{heal}. 
This leads to $N_{\rm c}\sim\ell^6/a {\cal L}^5$ in agreement with the theory of~\cite{gian09,gian09b}.}
In the Thomas-Fermi limit 
the disorder can be treated
perturbatively and a small negative linear shift in $\xi_{\rm heal}/{\mathcal L}$ would be expected~\cite{Falco07}. 
However, in figure~\ref{figureA}{a} a positive shift 
of the frequency of the breathing mode is found. This is not surprising since the present theory
has to be considered as exact only in the opposite limit $\xi_{\rm heal}\gg {\mathcal L}$
where disorder is dominating.

The frequencies plotted in figure~\ref{figureA}{a} refer to a single connected condensate.
However, above $N\geq{\mathcal L}/a$ the single connected condensate is a higher energy metastable state,
since the ground state 
is expected to undergo fragmentation~\cite{Nattermann08}. In the fragmented state, we do not expect
sharply defined frequencies. More likely, they should be distributed in some interval.
{\color{black} Note that the possibility to observe the single connected
condensate, in the regime where fragmentation is expected, could arise by means of 
a sudden decrease of the oscillator frequency. However, it is difficult to make further
estimates about the different timescales involved in the experiment.
In order to prevent fragmentation,} we would like to push the single fragment solution at larger particle numbers. 
For a given disorder (${\mathcal L}$) and trap configuration 
($\ell$) such that strong disorder condition $\ell\gg {\mathcal L}$
is satisfied, 
this can be achieved by maximizing the ratio 
${\mathcal L}/a$. Experimentally, this can be obtained 
by tuning the scattering length close to zero ({\it zero crossing}) by means of a Feshbach resonance
as proposed in Fig.~\ref{figureA}{b}.

\begin{figure}
\vspace{1cm}
\includegraphics[width=145mm,height=40mm]{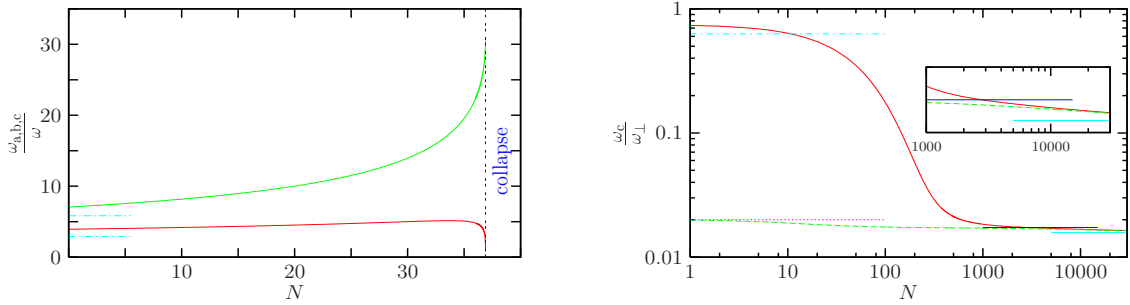}  
  \caption{(Colour online)(a) Example of the oscillations frequencies as function of the number of particles
  for negative scattering length. The upper solid (green) line describes the $a,b-$modes while the lower (red) 
  shows the $c-$mode. 
  The dashed (yellow) lines indicates the asymptotic 
  solutions~(\ref{eq:mode_b_strg_dis_b_0_smN})-(\ref{eq:mode_c_strg_dis_b_0_smN}).
  %The trap is isotropic
  %with  $\omega=2\pi 25$ Hz, $a=-100 a_0$ and ${\mathcal L}=200 a$
  %which implies
  %$\ell/{\mathcal L}\approx 2$. 
  (b)  
  Frequency of ``quadrupole" mode $c$ 
  (red solid line).
 % for a $^{87}$Rb condensate
 % in a strongly anisotropic cigar-shaped trap with $\omega_{\perp}=2\pi 150 $Hz,
 % $\omega_{\parallel}=0.01 \omega_{\perp}$, $a=150 a_0$,
 % and ${\mathcal L}=300 a$.
  The upper dot-dashed light blue line corresponds to 
  the analytic result of~(\ref{eq:mode_c_strg_dis_b_0_smN_oneD}).
  The solid dark-blue line indicates the frequency $\sqrt{3}~\omega_{\parallel}$
  characteristic of the $1D$ mean-field theory,
  while the solid light-blue represents the $\lambda\rightarrow 0$ limit of the $3D$ Thomas-Fermi frequency $\sqrt{5/2}~\omega_{\parallel}$.
  The clean theory is described by the dashed green line which at 
  low density approaches
  the free h.o. value $2\omega_{\parallel}$ (magenta dot-dashed). 
  The inset shows the details
  of the crossover from the $1D$ mean-field regime to the Thomas-Fermi condensate.   
} 
  \label{figure2}
\end{figure}

In the case of negative scattering length there is only a metastable state of
finite radius. This state becomes unstable 
at a critical particle number $N_c$.
%Interestingly, this value
%corresponds approximatively to the critical number of particles at which the {\it generalized} healing length $\xi$
%becomes equal to the size $R_0$ of the sample. This latter condition 
%can be considered as a rough estimation for the onset of superfluidity. 
In Fig.~\ref{figure2}a, we plot the frequencies of the three modes in the case
of an attractive interaction for an isotropic trap with  $\omega=2\pi 25$ Hz, $a=-100 a_0$ and ${\mathcal L}=200 a$.
%which implies
%$\ell/{\mathcal L}\approx 2$. 
The divergence signals the instability
of the gas when the number of particles reaches the critical value $N_c$.

\subsection{Anisotropic traps and lower dimensions}
\label{sec1c}

We have seen that, at low densities, when the physics %size of the system and the dynamics
is determined only by the balance between the random potential and the kinetic energy,
the frequencies of the oscillations
do not depend on the external harmonic potential. Therefore, the anisotropy of the trap becomes irrelevant.
Increasing the number of particles, 
the mean-field interaction becomes important and the 
harmonic trap affects indirectly the results through the Thomas-Fermi parameter $\alpha$. 
At even larger particle number the 
system enters the Thomas-Fermi regime where the harmonic trap dominates over the disorder and the anisotropy
plays its usual role \cite{Stringari96}.
If we consider, for example, an elongated cigar along the axial $z$-direction, with
$\omega_{x,y}\equiv\omega_{\perp}\gg \omega_{\parallel}\equiv\omega_z$, 
this description holds as long as we have ${\mathcal L}\ll \ell_{\perp}\ll \ell_{\parallel}$.
However, when $\ell_{\perp}\ll {\mathcal L}\ll \ell_{\parallel}$ the system can pass through the different regimes in the axial 
directions while its ground state in the radial direction is frozen to that of the harmonic oscillator of the radial
harmonic confinement.  %which In our case is expressed by the condition $r_{x0}=1$. 

Herewith, 
we limit our discussion to the single connected solution, neglecting the fragmented low-dimensional condensate 
where it occurs~\cite{Nattermann08}.
Using the definitions $a_{1D}=\ell_{\perp}^2/a$ and ${\mathcal L}_{1D}=({\mathcal L}\ell_{\perp}^2)^{1/3}$
we have that 
for small $N\ll  a_{1D}/
{\mathcal L}_{1D}$ the axial radius is determined by the equation $\gamma r_{z0}^{3/2}=1$.
This leads to the results 
\begin{eqnarray}
\label{eq:mode_b_strg_dis_b_0_smN_oneD}
&\omega_a=\omega_{b}=2\omega_{\perp}\\
\label{eq:mode_c_strg_dis_b_0_smN_oneD}
&\omega_{c}\propto \omega_{\perp}\left({\ell_{\perp}}/{\mathcal L}\right)^{2/3}.
\end{eqnarray}
At larger $N$ such $ a_{1D}/
{\mathcal L}_{1D} \ll  N \ll a_{1D}/{\mathcal L}_{1D}(\ell_{\parallel}/{\mathcal L}_{1D})^{4/5}$ 
%where $N_1\equiv a_{1D}/{\mathcal L}_{1D}(\ell_{\parallel}/{\mathcal L}_{1D})^{4/5}$ 
the axial radius is determined in the leading order by the equation $\gamma r_{z0}^{3/2}=\alpha r_{z0}$
and we have in first approximation 
\begin{eqnarray}
\label{eq:mode_a_strg_dis_b_0_TF_oneD}
&\omega_a=\omega_b=2\omega_{\perp}
\\
\label{eq:mode_c_strg_dis_b_0_TF_oneD}
&\omega_{c}\propto \omega_{\perp}\left({\ell_{\perp}}/{\mathcal L}\right)^{3/2}
\left({\ell_{\perp}}/{N a}\right)^{5/2}.
\end{eqnarray}
%Note, however, that for particular choice of the parameters in figure~\ref{figure2}b we have $N_1\approx 529$
%and then the condition $N_0\ll N\ll N_1$ cannot be fully satisfied. 
At even larger values the Thomas-Fermi regime is realized in the axial direction
while in the radial direction the ground state wave-function can still be determined 
by the free harmonic oscillator. Therefore, we have that the frequency of the breathing mode
approaches the value $\omega_c= \sqrt{3}\omega_{\parallel}$ characteristic
of the so-called $1D$ mean-field regime \cite{Stringari97}. In order to appear, this regime requires 
the two conditions $N a\lambda/\ell_{\perp}\ll 1 $ and $(N a/\sqrt{\lambda}\ell_{\perp})^{1/3}\gg 1$
\cite{Menotti02}.
Finally, for $N a\lambda/\ell_{\perp}\gg 1$ the gas enters the full Thomas-Fermi regime 
and the "quadrupole" mode approaches the strongly anisotropic $3D$ Thomas-Fermi result \cite{Stringari96}
$\omega_c=\sqrt{5/2}\omega_{\parallel}$ while $\omega_a=\sqrt{2}\omega_{\perp}$ and $\omega_b=2\omega_{\perp}$.
The full crossover is shown in Fig.~\ref{figure2}b for the 
lowest axial mode $c$, in a $^{87}$Rb condensate confined in a 
strongly anisotropic cigar-shaped trap with $\omega_{\perp}=2\pi 150 $Hz,
$\omega_{\parallel}=0.01 \omega_{\perp}$, $a=150 a_0$,
and ${\mathcal L}=300 a$. 
For these parameters,
$\ell_{\perp}/{\mathcal L}\approx 0.36$,
and $\ell_{\parallel}/{\mathcal L}\approx 3.7$. Moreover, 
the gas cannot enter the strong interacting Tonks regime
since it would 
require to fulfill the two conditions $\ell_{\perp}/a< 10^2$ and $N\lambda< 1$ simultaneously~\cite{Menotti02}.
In our case, we have $\sqrt{\lambda}\ell_{\perp}/a\approx 11$. Nevertheless, the Tonks gas
remains beyond the reach of our mean-field approach.

The treatment of a $2$D disc-like geometry trap when $\lambda\rightarrow\infty$
follows essentially the same outline. In this case we define
$\omega_{x,y}\equiv\omega_{\parallel}\ll \omega_{\perp}=\omega_z$.
Similar to the $1$D configuration, a non trivial interplay between disorder and low-dimensionality arises only when 
$\ell_{\perp}\ll {\mathcal L}\ll \ell_{\parallel}$. 
Except that for very large $N$, the ground state in $z$-radial
direction coincides with the lowest eigenstate of the harmonic trap, namely $r_{0z}^4=1/\lambda^2$.
The $2D$ analog of %the disorder dominated results of 
~(\ref{eq:mode_c_strg_dis_b_0_smN_oneD}) and~(\ref{eq:mode_c_strg_dis_b_0_TF_oneD})  
are respectively  $\omega_c\propto \omega_{\perp}{l_{\perp}}/{\mathcal L}$ in the small $N$ limit and 
$\omega_c\propto({l_{\perp}}/{\mathcal L})({l_{\perp}}/{N a})^{3/2}$ in the moderate interacting regime.
At larger values of $N$, for sufficiently strong anisotropies, 
the size of the cloud in the $x,y$ plane is determined by the competition between harmonic trap and interactions
while in the $z$-direction the ground state can still be that of the free harmonic oscillator. 
This is the $2D$ analog of the $1D$ mean-field regime for which $\omega_c=2 \omega_{\parallel}$ \cite{Ma00}.
Finally when $N\rightarrow \infty$ the $\lambda\rightarrow\infty$ limit of the $3D$ Thomas-Fermi 
is recovered where $\omega_c=\sqrt{10/3}~\omega_{\parallel}$ \cite{Stringari96}. 

\subsection{Finite correlation length} 
\label{sec1d}

The delta correlated disorder approximation
discussed so far, is appropriate  
for disorder with very short correlation length, {\it i.e.} 
when $b\leq {\mathcal{L}}$. 
In this case, the
correlation length of the disorder
is generally much shorter than the correlation length of the system, which
represents
a necessary condition for applying  
the Larkin scaling argument.
The small corrections due to finite $b$
can be taken into account using the full equations~(\ref{eq:equil_cond_x})-(\ref{eq:matrx_el_gen}).
Differently, when $b \gg {\mathcal{L}}$, 
the use of
the expression for disorder energy in~(\ref{eq:Var_Lagr})
is somehow questionable. 
{\color{black} Moreover, also self-averaging is expected to be much less efficient than in the limit delta correlated disorder.}
Therefore, we conclude that the description
based on~(\ref{eq:equil_cond_x})-(\ref{eq:matrx_el_gen}) 
holds
as far as the limiting case of delta correlated disorder is approached.

Nevertheless, it is interesting to make the following remark.
The classical limit $b \gg {\mathcal{L}}$ has been recently investigated in~\cite{gian09,Shklovskii08}.
Near one of its typical minima, the random potential can be approximated by a harmonic potential
well of depth $U_0=\kappa/b^{3/2}$ and width $b$. The bound state is located very close to the minima of the random potential
and the number of levels in a well is large.
From $U_0\equiv m \omega_b b^2/2$ we obtain the oscillator length of the 
typical fluctuating well as $\ell_b=b\left({\mathcal{L}}/b\right)^{1/8}$~\cite{gian09b,Shklovskii08}. Interestingly, this length can be obtained
from~(\ref{eq:equil_cond_x}) in the limit of small number of particle where the harmonic trap and the mean-field energy
can be neglected. At larger particle number the size of the fragment is determined by the competition of the mean-field interaction
and the disorder energy. This yields $R\sim(Na{\mathcal{L}}^{1/2} b^{7/2})^{1/5}$, which agrees with the size 
of the fragments in the multi-domain state found in~\cite{gian09b} by different methods.
This analysis reveals that the the Gaussian ansatz~(\ref{eq:Zoller_Ansatz})
is able to reproduce the ground-state localized state also for disorder with long-range
correlated potential.
%\noindent Next, we turn to consider how the dynamics of the collective oscillations in strongly disordered 
%traps is modified when the random potential has a finite correlation length $b$.
%Fig... shows the frequency of the quadrupole $c$-mode calculated substituting into
%the eigenvalues of the matrix...
%%Eqs.~(\ref{eq:genfreq})-(\ref{eq:gfunct}) 
%the equilibrium conditions  found 
%by solving numerically Eqs.~(\ref{eq:equil_cond_x}) and~(\ref{eq:equil_cond_z}). 
%We see that for a given value of the disorder strength a finite correlation length reduces the shift of the frequency due to the disorder
%and that the effect is stronger the larger the distance over which the disorder is correlated.
%In particular, the effects due to the random potential disappear at all when $b\gg \ell$.
%{\it What can be said about the self-averaging of the disorder induced frequencies shift when 
%$b$ becomes larger than the ``generalized'' healing length?}

%\noindent Alternatively, one can take 
%both the limits $b \rightarrow \infty$ and ${\mathcal L}\rightarrow 0$ simultaneously.
%Then the gas is trapped by the random potential which becomes harmonic with the effective
%frequency $\omega_{{b}_{\infty}}
%=(\hbar/m)\left(1/{\mathcal L~b}\right)^{1/4}$. In this special limit, the theory reduces to a that of 
%an harmonically trapped clean system with a renormalized Thomas-Fermi parameter   
%$\alpha'\equiv\alpha\sqrt{b^7 {\mathcal L}}/\ell^4$.

\begin{figure}
\vspace{1cm}
\includegraphics[width=150mm,height=40mm]{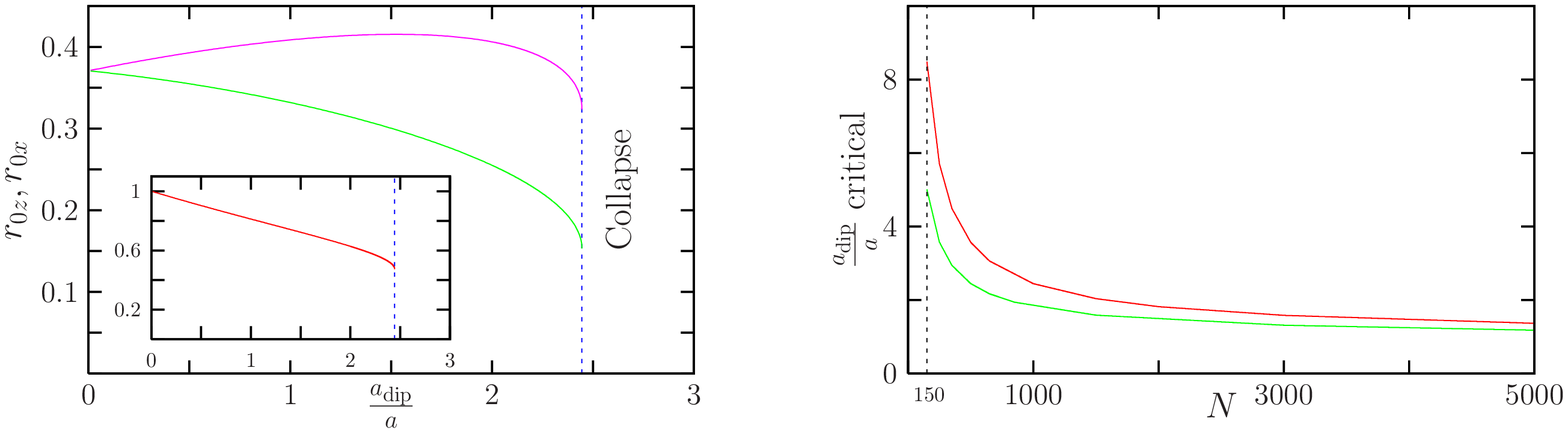}  
  \caption{(Colour online)(a) The size of the single fragment 
  of $^{52}$Cr atoms along the radial (green) and the axial (magenta) directions
  as a function of the ``relative" dipolar strength.
  The parameters are $N=1000$, $a=10 a_0$, $\lambda=1$, $\omega=2 \pi 10$ Hz, ${\mathcal L}\sim 2000 a$, and
  $\ell/{\mathcal L}\sim3.2$. 
  The inset shows their ratio $r_{0x}/r_{0z}$.
  (b) Dependence of critical value of the dipole interaction ($a_{\rm dip}/a$) from the particle number 
  for ${\mathcal L}\sim 2000 a$ (red) and for ${\mathcal L}\sim 1000 a$ (green).
  The other parameters are taken as in figure~\ref{figure3}(a). The dashed line at $N=150$ is just a guide for the eye.  
} 
  \label{figure3}
\end{figure}
\section{Dipolar gas}

\subsection{Stability of the gas}
\label{sec2a}

The time dependent variational approach can be extended to the presence 
of the long range anisotropic dipolar interaction~\cite{Yi01}. 
If we assume the dipoles aligned along the $z$-direction, 
the interaction between 
two polarized dipoles can be written as  
\begin{equation}
V\left(\bf r\right)
%\frac{d^2}{4\pi\epsilon_0}
%4\sqrt{\frac{\pi}{5}} \frac{Y_{20}(\theta)}{r^3},
%%\frac{1-3\cos{\theta}^2}{r^3}
=\frac{\mu_0\mu^2}{4\pi}4\sqrt{\frac{\pi}{5}} \frac{Y_{20}(\theta)}{r^3},
%\frac{1-3\cos{\theta}^2}{r^3},
\label{eq:dippot}
\end{equation}
where $\theta$ is the angle between ${\bf r}$
and the direction along which the dipoles are pointing.
The constant %$\epsilon_0$ 
$\mu_0$ is the magnetic constant and $\mu$ 
is the magnetic moment of the atoms.
The strength of the dipole interaction relative to the short range potential
will be expressed through the dimensionless quantity 
$\varepsilon
%\sqrt{\frac{\pi}{5}}\frac{d^2 m}{4 \pi^2 \epsilon_0\hbar^2 a}
%%(=\frac{\mu_0\mu^2 m}{12 \pi\hbar^2 a})
=\sqrt{{\pi}/{5}}\left(\mu_0\mu^2 m/4 \pi^2\hbar^2 a\right)$. 
%$\varepsilon=\frac{d^2 m}{12 \pi \epsilon_0\hbar^2 a}. 
Alternatively, the strength of the two
potentials can be compared by defining a characteristic length scale associate to the
dipole-dipole interaction. From the uncertainty principle, we can define  
$a_{\rm dip}$ as the distance at which the dipolar potential energy~(\ref{eq:dippot}) equals the kinetic
energy. This yields
\begin{equation}
a_{\rm dip}
%=\sqrt{\frac{\pi}{5}}\frac{d^2 m}{4 \pi^2 \epsilon_0\hbar^2 a},
=\sqrt{\frac{\pi}{5}}\frac{\mu_0\mu^2 m}{4 \pi^2 \hbar^2 },
\end{equation}
where we have chosen the prefactor such that $\varepsilon=a_{\rm dip}/a$ \cite{footnoteA}.

In presence of disorder and dipolar interaction, the equilibrium 
conditions of~(\ref{eq:equil_cond_x}-\ref{eq:equil_cond_z})
become
\begin{eqnarray}
\label{eq:equil_cond_x_dip}
r_{0x}^4+\gamma
\frac{r_{0x}}{\sqrt{r_{0z}}}
{f_{r_{0x}}}
{\prod_{j=1}^3 f_{r_{0j}}^{1/4}}
=1 + \alpha\frac{1}{r_{0z}}\left[1-\varepsilon {\mathcal F}\left(\frac{r_{0x}}{r_{0z}}\right)
%\left({\mathcal E}\right)
\right]\\
\label{eq:equil_cond_z_dip}
\lambda^2 r_{0z}^4+
\gamma\frac{r_{0z}^{3/2}}{r_{0x}}
{f_{r_{0z}}}
{\prod_{j=1}^3 f_{r_{0j}}^{1/4}}
=1+ \alpha\frac{ ~r_{0z}}{r_{0x}^2}\left[1-\varepsilon {\mathcal G}\left(\frac{r_{0x}}{r_{0z}}\right)\right].
% \label{eq:equil_cond}
\end{eqnarray}
with \cite{Yi01}
%\begin{eqnarray}
%{\mathcal F}\left(\xi\right)=\frac{\sqrt{5 \pi}}{6(1-\xi^2)^2}
%\left[
%-4\xi^4-7\xi^2+2+9\xi^4\frac{\rm{Arctanh}{\sqrt{1-\xi^2}}}{\sqrt{1-\xi^2}}
%\right],\nonumber
%\\
%{\mathcal G}\left(\xi\right)=\frac{\sqrt{5 \pi}}{3(1-\xi^2)^2}
%\left[
%-2\xi^4+10\xi^2+1-9\xi^2\frac{\rm{Arctanh}{\sqrt{1-\xi^2}}}{\sqrt{1-\xi^2}}
%\right].\nonumber
%\end{eqnarray}
${\mathcal F}\left(\xi\right)=\left[\sqrt{5 \pi}/6(1-\xi^2)^2\right]\left[
-4\xi^4-7\xi^2+2+9\xi^4{{\mathcal H}\left(\xi\right)}\right]$, and %{\rm{Arctanh}{\sqrt{1-\xi^2}}}/{\sqrt{1-\xi^2}}\right]$ and
${\mathcal G}\left(\xi\right)=\left[\sqrt{5 \pi}/3(1-\xi^2)^2\right]\left[
-2\xi^4+10\xi^2+1-9\xi^2{{\mathcal H}\left(\xi\right)}\right]$, %${\rm{Arctanh}{\sqrt{1-\xi^2}}}/{\sqrt{1-\xi^2}}
and ${\mathcal H}\left(\xi\right)={\rm{Arctanh}{\sqrt{1-\xi^2}}}/{\sqrt{1-\xi^2}}$.
\begin{figure}
\vspace{1cm}
\includegraphics[width=150mm,height=40mm]{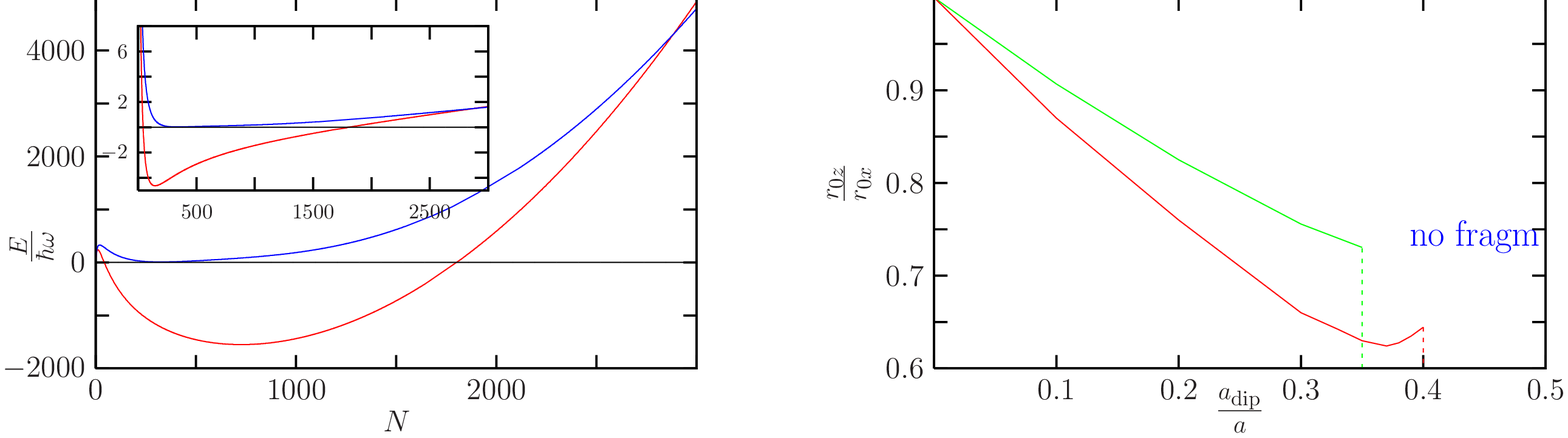}  
  \caption{(Colour online)(a) Energy of a $^{52}$Cr condensate
  in a strong random potential in absence of dipolar interaction (red)
  and for $a_{\rm dip}=0.40 a$ (blue) as function of $N$. 
  The other parameters are $a=100 a_0$, $\lambda=1$, $\omega=2 \pi 50$ Hz, ${\mathcal L}\sim 30 a$, and
  $\ell/{\mathcal L}\sim 9.5$. The inset shows the correspondent energy per particle. (b) 
  Typical deformation of the fragments as function of the relative dipolar strength for two different values
  of the disorder ${\mathcal L}\sim 50 a$ (green), ${\mathcal L}\sim 30 a$ (red). 
  The other parameters have been chosen as in figure~\ref{figure4}a.
  $^{52}$Cr atoms,
$a=100 a_0$, $\lambda=1$, $\omega=2 \pi 50$ Hz, upper ${\mathcal L}\sim 50 a$, lower ${\mathcal L}\sim 30 a$.} 
  \label{figure4}
\end{figure}
Analogously to the clean gas, these equations have at most only one stable solution. Here we are interested in the regime
of strong disorder and moderate interactions where the gas is confined mainly by the random potential
and is constituted by a single connected fragment. In this limit the anisotropy
of the harmonic trap does not play any role and, in absence of dipolar interaction, the cloud would have spherical symmetry. However,
turning on the dipolar potential the fragment tends to become more prolate along the direction of the dipoles orientation.
The situation is illustrated in figure~\ref{figure3}a, where we have considered a very short scattering length
in order to shift towards large $N$ the onset of multiple fragmentation which occurs above $N\geq {\mathcal L}/a$.
Moreover, such a small value of the scattering length, permits to 
access the strong dipolar regime $a_{\rm dip}/a\geq 1$ even for gases made of atoms with small magnetic moments.
This regime has been recently realized in experiments with $^{52}$Cr condensates 
near a Feshbach resonance \cite{Pfau05a,Pfau08_Nature}. For any fixed $N$, 
the fragment collapses at some critical value of $a_{\rm dip}/a$.
In figure~\ref{figure3}b we plot this threshold  
as function of the number of particles at fixed $a$ for two different values of the Larkin length.
For given $a$ and $N$ 
at shorter Larkin length the condensate is less stable against the attractive dipole interaction. 
Note that a shorter Larkin length amounts to a stronger center of attraction due to the 
fluctuations of the random potential and thus to a smaller volume occupied by the fragment.

\subsection{Fragmented state}
\label{sec2b}
So far we have not discussed the possibility of a multi-fragmented state since the variational method
is based on a single connected ground state. However, in~\cite{Nattermann08} it has been shown that
for a large region of the parameters space, strong disorder favours fragmentation.
It is therefore interesting to analyze 
the situation when a dipolar interaction is present as well.
In the spirit of~\cite{Nattermann08}, we evaluate the Hamiltonian operator
on the ground state~(\ref{eq:Zoller_Ansatz}).  
Assuming cylindrical symmetry and rescaling  energies in unities of 
the harmonic trap energy $\hbar\omega$,  
the energy per particle of the system can be written as
\begin{eqnarray}
\frac{\mathcal{E}}{\hbar\omega}=
%&
\frac{1}{2 r^2}
\left(
1
+\frac{\sigma^2}{2}
\right)+
\frac{1}{2} {r^2}\left(1+\frac{\lambda^2}{2\sigma^2}
\right)
%&
-\sqrt{\frac{\ell}{{\mathcal L}}}\sqrt{\frac{{\sigma}}{{r^3 }}}
+\frac{\alpha}{2}\frac{\sigma}{r^3}
\left[
1-\varepsilon I_1\left(\sigma
\right)
\right]\nonumber,
\end{eqnarray}
in terms of the variational parameters
$r\equiv r_{x}$ and $\sigma=r/r_z$.
The {\it deformation} function 
%$I_1$ 
$I_1(\xi)={\sqrt{5 \pi}}\left[
1+2 \xi^2-3\xi^2{{\mathcal H}\left(\xi\right)}
\right]/{3\left(1-\xi^2\right)}$
%\begin{eqnarray}
%I_1(\xi)=\frac{\sqrt{5 \pi}}{3\left(1-\xi^2\right)}
%\left(
%1+2 \xi^2-3\xi^2\frac{{\rm Arctanh}\sqrt{1-\xi^2}}{\sqrt{1-\xi^2}}
%\right)
%\end{eqnarray}
is continuous, monotonous and positive for $\xi<1$ and negative for $\xi>1$. 

In absence of dipolar interaction and for a strong given disorder potential the total energy $E(N)={N\mathcal E}$
has a minimum $E_1(N_1)$ at negative energy as shown in 
figure~\ref{figure4}a.
According to the Imry-Ma criterium~\cite{Nattermann08}, at $N$ larger than $N_1$
%the value where this minimum is reached, 
it is energetically favourable to split the gas into fragments
of energy $E_1$. Eventually, at even larger $N$,  
the oscillator energy per particle of the fragmented state 
%with size $R_{\rm F}$ 
reaches
the energy per particle from the disorder and the crossover to the Thomas-Fermi regime occurs \cite{Nattermann08}. 

The effect of the dipolar interaction on the phenomenon of the fragmentation can be seen as follows.
The attractive dipole interaction shrinks the total condensate volume causing
an increase of the kinetic energy which tends to delocalize the condensate
against the collective pinning induced by the disorder. This is consistent with the numerical results shown in 
figure~\ref{figure4}a.
At fixed disorder, finite values of the dipolar interaction shift upward
the minimum of the energy $E={N\mathcal E}$ until it becomes 
positive. Then, strictly speaking, the possibility
of having fragmentation is ruled out at least at the level of the Imry-Ma argument. 
Nevertheless, at sufficiently small dipolar interaction fragmentation can occur. 
In the presence of anisotropic interactions the Imry-Ma \cite{Imry75} argument 
does not imply that domains have to be spherical. 
The typical fragment will be preferentially cigar-shaped with the long axis parallel to the dipoles
\cite{Nattermann88}.
%In general domains
%must not be spherical and the dipolar interaction will favour cigar-shaped ellipsoids with the long axis
%parallel to the dipoles \cite{Nattermann88}.
This means that, while in a single realization of the disorder the various fragments should appear with random shape,
they will have on average a characteristic deformation.
The dependence on the dipolar interaction of the typical deformation 
can be estimated evaluating the equilibrium radii $r_{0x}$ ad $r_{0z}$ 
in the minimum of the energy curve $E(N)$.
The result obtained is shown in  
figure~\ref{figure4}b for two different realization 
of the disorder. The curves stop where the Imry-Ma argument for the existence of fragmentation becomes invalid. 
Note that, for the chosen parameters, this boundary
is smaller than the critical dipolar strength at which the cloud would collapse.

\section*{Conclusions}

We have proposed a hybrid approach, originated from the variational method for the Gross-Pitaevskii theory
and the Larkin-Imry-Ma scaling argument of~\cite{Nattermann08}, aimed to study zero temperature Bose-Einstein condensates
in strongly disordered traps. 
{\color{black} Similarly as for other disordered systems, 
the Larkin-Imry-Ma scaling analysis, despite its intrinsic limitations,
allows to make semi-quantitative predictions in situations where 
no mean-field description seems possible~\cite{Hertz85}.}
The theory addresses the problem 
of the condensate in the limit of strong disorder and moderate interaction.
There, the dynamics of the condensate is dominated by a new length scale, namely the Larkin length, 
associated to the collective pinning of the condensate.
{\color{black} In the case of isotropic interaction, we have investigated the stability and the low-lying excitations
of the system
at different dimensionalities when the ground state consists of a single connected condensate. We have suggested that, 
for given disorder strength and trapping  parameters, this regime could be observed experimentally
by using scattering length zero crossing \cite{Fattori08}.
%tuning the scattering length near to zero using Feshbach resonance techniques. 
In the case of dipolar interactions, our theory offers in addition the possibility to investigate
the effects of
the anisotropic interaction on the phenomenon of fragmentation.
Predictions about the typical deformation ratio of the fragments can be made.
We conclude that,} the realization of types of disorder with correlation length
smaller than all other length scales, could allow to test the theory in the laboratory.

\ack{I would like to thank T. Nattermann and V. L. Pokrovsky for useful discussions.}
%\section*{Acknowledgements}
%I would like to thank A. Pelster, T. Nattermann and V. L. Pokrovsky for useful discussions.


\begin{thebibliography}{99}
\bibitem{heal}
In the non superfluid regime we adopt here the definition of the {\it generalized}
healing lenght $\xi_{\rm heal}$~\cite{Nattermann08}. The latter is defined as the length for which
$\hbar^2/m\xi_{\rm heal}^2\sim \hbar^2 N a/m R_0^3$ holds where $R_0$ is the size of the system at equilibrium.
{\color{black} In the superfluid regime the generalized healing length coincides with the Thomas-Fermi
healing length $\xi_{\rm heal}=\ell (N a/\ell)^{-1/5}$.}
\bibitem{frgm}
In the fragmented state, we expect
the frequencies to be not sharply defined but to be distributed in some interval.
\bibitem{mettinota}
In this paper we neglect systematically the logarithmic corrections found recently
in~\cite{gian09}.
\end{thebibliography}
\end{document}